\documentclass[10pt]{article}

\usepackage{amsmath}
\usepackage{amssymb}
\usepackage{graphicx}
\usepackage{verbatim}
\usepackage{subfigure}

\newcommand{\be}{\begin{equation}}
\newcommand{\ee}{\end{equation}}
\newcommand{\bea}{\begin{eqnarray}}
\newcommand{\eea}{\end{eqnarray}}

\newcommand{\rc}{\nonumber\\}

\numberwithin{equation}{section}

\begin{document}

%%%%%%%%%%%%%%%%%%%%%%%%%%%%%%%%%%%%%%%%%%%%%%%%%%%%%%%%%%%%%%%%%%%
%\begin{flushright}
%ECM-XXXX\\
\rightline{ICCUB-11-158}

\rightline{July, 2011}
%\end{flushright}

\bigskip

\begin{center}

{\Large\bf  
 ${\cal N}=1$ SQCD-like theories with $N_f$ massive flavors 
\\
from AdS/CFT and $\beta $ functions
}
\bigskip
\bigskip

{\large Alejandro Barranco \footnotemark[1], Elisabetta Pallante \footnotemark[2]  and Jorge G. Russo  \footnotemark[1] \footnotemark[3]}
\bigskip

{\it
1) Institute of Cosmos Sciences and Estructura i Constituents de la Materia\\
Facultat de F{\'\i}sica, Universitat de Barcelona\\
Av. Diagonal 647,  08028 Barcelona, Spain\\
\smallskip
2) Centre for Theoretical Physics, University of Groningen, 9747 AG, Netherlands\\
\smallskip
3) Instituci\'o Catalana de Recerca i Estudis Avan\c cats (ICREA)\\
Pg. Lluis Companys, 23, 08010 Barcelona, Spain\\
}
\bigskip
\bigskip

\end{center}
\bigskip

\begin{abstract}

We study new supergravity  solutions related to large-$N_c$  ${\cal N}=1$ supersymmetric gauge field theories with a  large number $N_f$  of massive flavors. We use a recently proposed framework based on configurations with $N_c$ color D5 branes and a distribution of $N_f$ flavor D5 branes, governed by a function $N_f S(r)$. Although the system admits many solutions, under plausible physical assumptions
the  relevant solution is   uniquely determined for each value of $x\equiv N_f/N_c$.
In the IR region, the solution smoothly approaches the deformed Maldacena-N\'u\~nez solution. In the UV region it approaches
a linear dilaton solution. 
For  $x<2$ the gauge coupling $\beta_g $ function computed holographically is negative definite,  in the UV
approaching the NSVZ $\beta $ function  with anomalous dimension $\gamma_0= -1/2$ (approaching $-3/(32\pi^2)(2N_c-N_f)g^3$)), and with $\beta_g \to-\infty $ in the IR.
For $x=2$, $\beta_g$ has a UV fixed point at strong coupling, suggesting  the existence of an IR fixed point at a lower value of the coupling.
We argue that the solutions with $x>2$  describe a ``Seiberg dual" picture where $N_f-2N_c$ flips sign.

\end{abstract}

\clearpage

\tableofcontents
%%%%%%%%%%%%%%%%%%%%%%%%%%%%%%%%%%%%%%%%%%%
\textwidth 16cm
\oddsidemargin .5cm
%%%%%%%%%%%%%%%%%%%%%%%%%%%%%%%%%%%%%%%

\section{Introduction}
\setcounter{equation}{0}

An important area of application of AdS/CFT dualities concerns ${\cal N}=1$ supersymmetric
$SU(N_c)$ Yang-Mills theory with an arbitrary number $N_f$ of fundamental flavors. 
The $N_f=0$ case   was constructed by Maldacena and N\' u\~nez (MN) \cite{Maldacena:2000yy}, building up on a geometry previously found in \cite{Chamseddine:1997nm}.
Massless flavors in the fundamental representation of the $SU(N_c)$ gauge theory can be incorporated  following the idea of \cite{Karch:2002sh} by adding $N_f$ spacetime filling branes.  
%interesting limits are the quenched limit $N_f\to 0$ and the Veneziano limit, with $N_f/N_c$ fixed for $N_c\to\infty$.
The resulting holographic models   \cite{Casero:2006pt,Casero:2007jj,HoyosBadajoz:2008fw} have led to many interesting physical insights, including, for instance, aspects of Seiberg duality (see also \cite{Paredes:2006wb,Bertoldi:2007sf,Caceres:2007mu}).
However, the presence of a singularity in the IR region limits the applicability of this geometry.
Recently, a new  ${\cal N}=1$ supersymmetric geometry has been found by Conde, Gaillard and Ramallo \cite{Ramallo} which includes the previous ones as particular cases, but more generally can
circumvent the IR singularity. The solution is parametrized in terms of a function $S(r)$ that encodes how flavor branes
are distributed in the space. The plan of this paper is to use this framework  to construct new solutions and
investigate new physical properties, with the aim of understanding the extent to which these geometries can describe aspects of 
${\cal N}=1$ supersymmetric
$SU(N_c)$ Yang-Mills theory with  $N_f$  fundamental massive flavors.

${\cal N}=1$ supersymmetric
$SU(N_c)$ Yang-Mills theories with  $N_f$  massless flavors, analogously to their non-supersymmetric counterpart, are likely to abandon the QCD-like confined phase for sufficiently large number of flavors and develop a conformal phase before the loss of asymptotic freedom. The restoration of conformal symmetry and 
the presence of a so called  conformal window in the number of flavors would thus identify a new family of non abelian gauge theories which is worth to explore. From the more phenomenological point of view, conformal symmetry might play a relevant role in the description of particle dynamics at energies above the electroweak symmetry breaking scale. 
\medskip

\noindent 
Until now, the emergence of conformal symmetry in theories without supersymmetry has been discussed in the context of Schwinger-Dyson equations for chiral symmetry in the ladder approximation
\cite{Appelquist_ladder}, truncated non-perturbative RG flows \cite{BraunGies}, supersymmetry inspired conjectures \cite{Sannino_Susy} and deformation theory \cite{Poppitz}. 
The proof of existence of a conformal window however depends on our ability to describe these theories in a non-perturbative manner, following the evolution of parameters all the way from strong coupling to weak coupling. Lattice studies are currently the only ones to provide a fully non perturbative analysis, and in $N_c=3$ QCD they have recently produced evidence that $N_f=12$ is plausibly close to the end-point of a conformal window \cite{Appelquist_Lat,Hasenfratz,Deuzeman,kuti}. Similar results have been found in \cite{Armoni:2009jn} using the worldline formalism.

\medskip

\noindent A further insight comes from supersymmetric gauge field theories.
 The renormalization group of ${\cal N}=1$ supersymmetric QCD (SQCD) has been extensively studied, and the perturbative $\beta$ function for the gauge coupling
is given by the well known NSVZ  formula \cite{Novikov:1983uc} 
\be
\beta_g = -{g^3\over 16\pi^2}\ {3N_c-N_f(1-\gamma_0)\over 1-{g^2 N_c\over 8\pi^2}}\, ,
\label{betagg}
\ee
where $\gamma_0$ is the anomalous dimension of the $N_f$ (massless) fundamental  superfields $Q^r $ and $\tilde Q_u$. 
The formula (\ref{betagg}) is unchanged, with a common anomalous dimension for all fundamental fields,  as long as the superpotential preserves a  $Z_{N_f}$ flavor symmetry. It also applies in the massive flavor case in the far ultraviolet, whereas it is drastically modified below the mass scale of the $N_f$ flavors.
 The anomalous dimension $\gamma_0$ is a non-trivial function of $N_f,\ N_c$, the gauge  coupling and all other couplings appearing in the superpotential. All the information about the dynamics of the strongly or weakly interacting theory is thus contained in $\gamma_0$. Its form depends on the precise choice of the superpotential  and determines the presence of fixed points in the parameter space of the theory. Any rigorous prediction for the existence and width of a conformal window would thus require to derive the $\beta$ functions and anomalous dimensions of the theory in a non-perturbative way. 
Holographic techniques may  allow to study the renormalization group flow beyond perturbation theory.
In particular, for the supersymmetric pure Yang-Mills theory (SYM), i.e. SQCD with $N_f=0$, the gravity dual was constructed in \cite{Maldacena:2000yy} and reproduced some interesting features of
${\cal N}=1$ SYM, including the correct structure of the  NSVZ $\beta $ function (\ref{betagg}) at $N_f=0$ for the gauge coupling \cite{DiVecchia:2002ks,Bertolini:2002yr}. 
The holographic $\beta $ function contains,  in addition,  non-perturbative contributions, and the gauge coupling 
goes to infinity in the IR -- a realization of ordinary confinement.

\medskip

\noindent The holographic approach has some well known limitations that we will not attempt to resolve in this paper.
One limitation is that holographic models typically include an infinite tower of 
 Kaluza-Klein states which have no counterpart in the field theory that one would like to describe, e.g. super QCD or super YM, and should therefore be decoupled. It was argued in \cite{Casero:2007jj,HoyosBadajoz:2008fw} that integrating out Kaluza-Klein excitations  leads to effective
couplings -- e.g. quartic couplings -- in the superpotential, so that the final effective field theory will be  an ${\cal N}=1$ supersymmetric QCD with the addition of an effective superpotential containing higher dimensional couplings. 
Such couplings can lead to the disappearance of the conformal window.
An example is the role of four-fermion operators in non-supersymmetric gauge theories. 
The presence of such operators has to be traced back to chiral dynamics and the breaking of chiral symmetry, with consequent disappearance of conformality.\footnote{ 
In particular, the Schwinger-Dyson gap equation in the ladder approximation says that the onset of chiral symmetry breaking occurs for a critical gauge coupling where the anomalous dimension of the fermion mass operator $\gamma_m=1$ and thus the four-fermion operator becomes relevant in the RG sense. In \cite{Cohen} has been also suggested that, while the value of the critical coupling will be affected by higher order effects in the perturbative (ladder) expansion, the scaling of the fermion propagator with $\gamma_m=1$ is the non-perturbative signal of chiral symmetry breaking.}   

\medskip

\noindent Despite these limitations of the holographic approach, it remains important to investigate the holographic predictions for the gauge coupling $\beta$-function, since they provide  detailed hints on the possible structure of the effective field theory in the strong coupling regime.
Such hints may be helpful  when studying the theory on the lattice or by means of other techniques.
In particular, for $N_f=2N_c$, a  prediction arising from our study is the existence of a non-trivial UV fixed point at some strong coupling $g_*$. 
Consistency with the RG evolution at weak coupling requires the existence of an IR fixed point at $ g_*'< g_*$, as we shall discuss.
Notice that the presence of a UV fixed point at strong coupling in addition to an IR fixed point has been conjectured already in the pioneering work by Banks-Zaks \cite{BanksZaks}, and might lead to a mechanism of disappearance of the conformal window via the annihilation of a pair of fixed points as suggested in \cite{Kaplan:2009kr}.
\medskip

\noindent This paper is organized as follows. 
In section 2 we review the type IIB string  backgrounds recently constructed in \cite{Ramallo}.
The solutions are parametrized by a profile  function $S(r)$ which for particular choices reproduces previous solutions in the literature.  
In section 3, we will adopt a convenient choice of $S(r)$ and obtain numerical solutions  for the three cases: i) $N_f<2N_c$, ii) $N_f =2N_c$ and iii) $N_f>2N_c$. The gauge coupling $\beta$-functions of the corresponding gauge theory for all cases are derived in section 4, with focus on the emergence of fixed points in the RG flow. 
We conclude in section 5, comparing our results with physical expectations and discussing possible future directions.

%%
%%%%%%%%%%%%%%%%%%%%%%%%%%%%%%%%%%
\section{The string/supergravity background}
%%%%%%%%%%%%%%%%%%%%%%%%%%%%%%%%%%
\setcounter{equation}{0}
We consider a type IIB supergravity  background with $N_f$ D5  branes extended on a non-compact two-cycle of a CY3-fold.  The ansatz is given by the following (Einstein frame) metric and RR three-form \cite{Ramallo}
\bea
ds^2 &=& e^{2f(r)}
 \Big[dx_{1,3}^2 + e^{2k(r)}dr^2
+ e^{2 h(r)}
(d\theta^2 + \sin^2\theta d\varphi^2) \nonumber\\
&+&\frac{e^{2 g(r)}}{4}
\left((\tilde{\omega}_1+a(r)d\theta)^2
+ (\tilde{\omega}_2-a(r)\sin\theta d\varphi)^2\right)
 + \frac{e^{2 k(r)}}{4}
(\tilde{\omega}_3 + \cos\theta d\varphi)^2\Big], \nonumber\\
F_{(3)} &=&\frac{N_c}{4}\Bigg[-(\tilde{\omega}_1+b(r) d\theta)\wedge
(\tilde{\omega}_2-b(r) \sin\theta d\varphi)\wedge
(\tilde{\omega}_3 + \cos\theta d\varphi)\nonumber\\
&+& (b'(r)+L_1(r))dr \wedge (-d\theta \wedge \tilde{\omega}_1  +
\sin\theta d\varphi
\wedge
\tilde{\omega}_2) + (1-b(r)^2+L_2(r)) \sin\theta d\theta\wedge d\varphi \wedge
\tilde{\omega}_3\Bigg]\nonumber\\
&-&\frac{N_c}{2}\sin\theta d\theta \wedge
d\varphi \wedge 
\tilde{\omega}_3\ ,
\label{nonabmetric424}
\eea
where  the $\tilde\omega_i$ are the left-invariant forms of $SU(2)$ given by
\bea\label{su2}
&&\tilde{\omega}_1= \cos\psi d\tilde\theta\,+\,\sin\psi\sin\tilde\theta
d\tilde\varphi\ ,\rc
&&\tilde{\omega}_2=-\sin\psi d\tilde\theta\,+\,\cos\psi\sin\tilde\theta
d\tilde\varphi\ ,\rc
&&\tilde{\omega}_3=d\psi\,+\,\cos\tilde\theta d\tilde\varphi\ ,
\eea
and  $a,b,f,k,g,h$ are  functions of the radius $r$.  This background embeds a flavor deformation by means of  a stack of flavor D5 branes parametrized by the $L_i(r), i=1,2$ functions.   It is convenient to introduce the functions $P$, $Q$ and the profile function $S$
\be
Q(r)\equiv e^{2g} (a \cosh(2r)-1)\ ,\qquad P(r)\equiv ae^{2g}\sinh(2r)\ ,\qquad S(r)\equiv-\frac{N_c}{N_f}L_2(r)\, .
\label{eq:PQS}
\ee 
The inverse relations for the metric functions $g(r)$ and $a(r)$ read
\be
e^{2g}= P\ \coth(2r)-Q\ ,\qquad a=\frac{P}{P\cosh(2r)-Q\sinh(2r)}\ .
\ee
The profile $S(r)$ characterizes the embedding of flavor D5 branes and it may be interpreted as an energy-scale dependence of the effective flavor number $N_fS(r)$ in the description of fundamental massive flavors of the dual gauge theory. The BPS equations can be solved in terms of the functions in (\ref{eq:PQS}) as follows
\begin{align}
L_1(r)&=\frac{N_f}{N_c}\frac{S'(r)}{2\cosh(2r)}\ ,\\
b(r)&=\frac{2r}{\sinh(2r)}-\frac{N_f}{N_c}\frac{S(r)}{2\cosh(2r)}+\frac{2}{\sinh(2r)}\int_0^r\tanh(2\rho )S(\rho )d\rho\
,\\
&\\
\label{hhj}
e^{2h}&=\frac{1}{4}\frac{P^2-Q^2}{P\coth(2r)-Q}\ ,\hspace{25mm} e^{2k}=\frac{P'+N_fS(r)}{2}\
,\\
f&=\frac{\Phi}{4}\ ,\hspace{45.75mm} e^{-2\Phi}=e^{-\Phi_0}\frac{e^{h+g+k}}{\sinh(2r)}\
,
\end{align}
where $\Phi (r)$ is the dilaton and $\Phi_0$ a constant. The function $Q$ is an integral over the profile $S(r)$ 
\be\label{Q}
Q=\coth(2r)\left[\int_0^r\frac{2N_c-N_fS(\rho )}{\coth^2(2\rho )}d\rho +q_0\right]\, ,
\ee
with $q_0$ a constant of integration. Finally, the function $P(r)$ satisfies the ``master" differential equation
\be\label{mastereq}
P''+N_fS'+(P'+N_fS)\left(\frac{P'-Q'+2N_fS}{P+Q}+\frac{P'+Q'+2N_fS}{P-Q}-4\coth(2r) \right)=0\ .
\ee
Once the profile $S(r)$ is determined,  a solution is obtained by first computing $Q(r)$ in (\ref{Q}), then solving (\ref{mastereq}) for $P(r)$. 
It should be noted that regularity of the geometry (see (\ref{hhj})) requires 
\be\label{reg. conds.}
P  >\vert Q\vert\ ,\qquad P'>-N_fS\ .
\ee
%%
%The role of the function $S(r)$ is the one of describing an ``effective" number of flavors $N_fS(r)$ which varies with the radius of the string dual background from zero to $N_f$, for $S=1$. 
%
In the particular cases $S=0$ and $S=1$ one finds solutions which already appeared in the literature.
We briefly review these cases before moving to the massive case described by the function $S(r)$.

\subsubsection*{$S=0$: Unflavored solution}\label{sectionS=0}

\noindent Setting $S=0$ in \eqref{mastereq} amounts to set $N_f=0$. 
In addition to the regular solution of the Maldacena-N\' u\~nez model \cite{Maldacena:2000yy,Chamseddine:1997nm},
there is a one-parameter deformation found in  \cite{Casero:2006pt} that leads to solutions with regular behavior at $r=0$.
  The infrared asymptotic of this unflavored one-parameter family of solutions has been explicitly written for $P$ in \cite{HoyosBadajoz:2008fw},
\be\label{IRMMN}
P=h_1r+\frac{4h_1}{15}\left(1-\frac{4N_c^2}{h_1^2}\right)r^3+\frac{16h_1}{525}\left(1-\frac{4N_c^2}{3h_1^2}-\frac{32N_c^4}{3h_1^4}\right)r^5+\mathcal O(r^7)\ ,
\ee 
where $h_1$ is the one parameter that labels each solution of the family. When $h_1=2N_c$, one recovers the
Maldacena--N\' u\~nez (MN) solution \cite{Maldacena:2000yy}. It is worth  noting that the resulting function $Q$ is the same for any value of $h_1$,
\be
Q=N_c(2r\coth(2r) - 1)\ .
\ee

\subsubsection*{$S=1$: Massless flavors}\label{sectionS=1}
By setting $S=1$ one obtains the  solutions of \cite{Casero:2006pt} describing massless flavors.
The asymptotic for these solutions was discussed in full detail in \cite{HoyosBadajoz:2008fw}. 
For large radius, i.e. in the ultraviolet of the dual gauge theory, a generic solution behaves exponentially 
\be
P= k\ e^{4r/3} + \mathcal O(1)\ ,
\ee
where $k$ is an integration constant.
There are also special solutions with the following linearly rising large $r$ asymptotic:
\be
P=  |2N_c-N_f| \ r + \mathcal O(1)\ ,\qquad N_f\neq 2N_c \ .
\label{linearP}
\ee
A further analysis is required to see if the geometry can actually be
extended to $r\to\infty $ or if it meets a singularity before. This will be discussed below. When  $N_f=2N_c$, there are special solutions
with the following asymptotic behavior:
\be\label{asymptoticS=1}
P=P_0+\mathcal O( e^{-c r}) \ ,\qquad P_0 = \frac{8N_c}{\xi(4-\xi)}\ ,
\ee
with
\be
q_0 = {4N_c (\xi-2)\over \xi(4-\xi)} \ ,\qquad c=1+\sqrt{9-4\xi+\xi^2}\ ,\qquad 0<\xi<4 \ .
\ee

\subsubsection*{$S(r)$: Massive flavors}
\label{sec:massive}

Following \cite{Ramallo} one can characterize the supersymmetric D5 brane embeddings by two algebraic equations
\be
F_1(z_i)=0\ ,\qquad F_2(z_i)=0\ ,
\ee
for the four complex variables $z_i$ ($i=1,\,\ldots\,,4$) parametrizing a deformed conifold,  thus satisfying
\be
\label{eq:conifold}
z_1z_2-z_3z_4=1 
\ee  
and related to the radial coordinate through 
\be
\label{eq:zi_radius}
\sum_{i=1}^4\vert z_i\vert^2=2\cosh(2r)\ .
\ee
In particular, the choice made in \cite{Ramallo} is given by the following embedding parametrized by two complex constants $A$ and $B$:
\be\label{embedding}
z_3=Az_1\ ,\qquad z_4=Bz_2\ .
\ee
This equation, together with \eqref{eq:conifold} and \eqref{eq:zi_radius} determines the minimum distance $r_q$ that this embedding reaches
\be
\cosh(2r_q)=\frac{\sqrt{1+\vert A\vert^2}\sqrt{1+\vert B\vert^2} }{\vert 1-AB\vert }\, .
\label{eq:rq}
\ee 
It depends on the moduli of $A$ and $B$, as well as their phase. 
By demanding that  the WZ term of the action of the full set of D5 branes in the ten-dimensional theory
coincides with the action obtained from  the embeddings
one arrives at 
%the following equation  
\cite{Ramallo}
%\be
%e^{2k}S+\frac{1}{2}e^{2g}\tanh(2r)S'=\mathcal S(r)\ ,
%\label{eq:profile}
%\ee
%where $\mathcal S(r)$ is related to the pullback of the K\"ahler form $J$ through
%\bea
%\int_{\mathcal C_2}\iota^*(J) &=& 2\pi\int dre^{-\frac{\Phi}{2}}\mathcal S(r)\
%,\\
%e^{-\frac{\Phi}{2}}J&=&\frac{e^{2k}}{2}dr\wedge(\tilde \omega_3+\cos\theta d\varphi)^+\frac{e^{2g}}{4}\frac{a\cosh(2r)-1}{\sinh(2r)}(d\theta\wedge\tilde\omega_2+\sin\theta d\varphi\wedge\tilde\omega_1)\\&&-\frac{e^{2g}}{4}\left(\frac{a\cosh(4r)-\cosh(2r)}{\sinh(2r)}\sin\theta d\theta\wedge d\varphi+\frac{\cosh(2r)-a}{\sinh(2r)}\sin\tilde\theta d\tilde\theta\wedge d\tilde\varphi\right)\, .
%\eea
%Applying this relation to the embedding \eqref{embedding} one obtains the function 
%%
%\be
%{\mathcal S(r)} = (6.16)\, .
%\ee
%%
%and the solution of \eqref{eq:profile} for the given embedding provides the profile function
\be\label{S(r)}
S(r)=\frac{\sqrt{\cosh 4r-\cosh 4r_q}}{\sqrt{2}\sinh(2r)}\Theta(r-r_q)\ .
\ee
%where $r_q$ is the minimum distance that the embedding reaches. 
Notice that $S(r)$ is continuous at $r=r_q$, while $S'(r)$ diverges as $S'(r)\sim(r-r_q)^{-1/2}$ near $r_q$, and it is thus singular.
To avoid this singularity Conde, Gaillard and Ramallo \cite{Ramallo} have proposed a brane setup for which the tip of the branes $r_q$ 
 is ``smeared", so that an average should be made over  brane distributions with different tip positions, weighted with a density function $\rho(r_q)$.
After performing the change of variables $y=\cosh(4r)$ and $y_q=\cosh(4r_q)$ with $y\geq 1$, and assuming that  the branes are distributed
over the whole space $0<r<\infty$,  then  the profile function will be given by
\be\label{S-rho}
S(y)=\int_{1}^{y}dy_q \ \rho(y_q)\frac{\sqrt{y-y_q}}{\sqrt{y-1}}\, ,
\ee 
where the measure function $\rho(y_q)$ satisfies the normalization condition
\be
\int_1^\infty \ dy_q\rho(y_q)=1\ .
\label{norma}
\ee

\section{Simple solutions for massive flavors}
\setcounter{equation}{0}

On the gauge field theory side, one expects that 
the asymptotic physics for $N_f$ massive flavors at high and low energies should be as follows:
a)  at energies lower than the flavor mass (infrared limit) it should converge to the unflavored case, $S=0$; b)  at  high energies (ultraviolet limit)
it should converge to the $N_f$ massless flavors case, $S=1$. 
This picture can be realized by the gravity dual background when the function $S(r)$ interpolates between the infrared/small radius limit $S(r)\to 0$ for $r\ll r_q$ -- with  $r_q$  being a measure of the common quark mass -- and the ultraviolet/large radius limit $S(r)\to 1$ for $r\gg r_q$.

\noindent Thus we are interested in solutions for massive flavors that approach the deformed MN solution (\ref{IRMMN}) in the infrared, i.e. in the small radius limit $r\to 0$. 
%(CHECK $r\to r_q$).
In \cite{Ramallo}, to describe flavors with a given  mass $\mathcal
O(y_q) $ with some spread, a measure function $\rho(y_q)$ with a finite
support around $y_q$ was chosen.
 Here, we slightly depart from this approach. Given the freedom in the choice of distribution of branes, we will conveniently adopt a smooth distribution $\rho(y_q)$ of branes, chosen to meet the following requirements:

\begin{itemize}

\item $S(r)$ is assumed to be a monotonous, continuous function varying between 0 and 1, approaching 1 at infinity.
We demand $S(r)\sim r^4$ (or smaller)  for $r\sim 0$, so that the curvature invariants of the geometry  near $r=0$ are the same
as in the deformed MN solution. In this way we ensure that the metric is regular at the origin (and  that there are no massless flavors).

\item In order to have a more tractable differential equation \eqref{mastereq}, we demand that $S$ is such that the integral (\ref{Q}) defining  $Q$ can be explicitly performed
with a simple result for $Q$.

\item Finally, we demand that $\rho(y_q)$ is positive definite and satisfies the normalization condition \eqref{norma}.
%  and that has a simple expression.

\end{itemize}

\medskip

\noindent We found  an extremely simple choice that meets all these requirements:
\be
S(r)=\left(\tanh(2r)\right)^{2n}\ ,\hspace{2cm} n=2,\,3,\,\dots%n=1
\label{aspa}
\ee
This corresponds to a distribution of branes with masses concentrated around the maximum of $S'(r)$,
at
\be
r_{\rm max}={\rm arccoth}\left(\sqrt{\frac{3+2 n+2 \sqrt{4 n+2}}{2 n-1}}\right)\ ,
\label{rmx}
\ee
which increases with $n$ (for large $n$, $r_{\rm max}\sim 1/4 \log n $). The spread $\Delta r$  decreases with $n$.
In order to determine $\rho(y_q)$, we note that the integral defining $S$ is related to an Abel Transform as follows
\be
2\partial_y(\sqrt{y-1}\ S(y))= {\mathcal A}[ \rho(y)] = \int_1^y dy_q \ \frac{\rho (y_q)}{\sqrt{y-y_q}}\
.
\ee
The inverse Abel Transform formula is
\be
\rho(y_q)=\frac{2}{\pi} \partial_{y_q}\int_1^{y_q}\frac{\partial_y\left(\sqrt{y-1}\ S(y)\right)}{(y_q-y)^{1/2}} dy\
.
\ee
It is easy to verify that the normalization condition (\ref{norma}) is satisfied for this measure function.

\noindent For the choice (\ref{aspa}), we  find 
\be
\rho_{(n)}(y_q) = \frac {4\sqrt{2}\ \Gamma(n+{3\over 2}) } {\sqrt{\pi} (n-1)! }\ \frac{ (y_q-1)^{n-1}}{(y_q + 1)^{n+{3\over 2}}}\ .
\ee
In particular
\bea
%\rho_{(n=1)}(y_q) &=& \frac{ 3\sqrt{2}}{(1+y_q)^{5\over 2}}\ ,
%\NO\\
%\rho_{(n=3)}(t) &=& \frac{ 16\sqrt{2(t-1)}}{\pi (1+t)^(3)}   
%\NO\\
\rho_{(n=2)}(y_q) &=& \frac{ 15 (y_q-1)}{\sqrt{2}  (1+y_q)^{7\over 2}}\ . 
\eea
%$\rho_{(n=2)}(y_q) $ has a maximum at $y_q=4/3$.
Next, we compute $Q(r)$ in (\ref{Q}).
The basic integral we need is
\be
\int_0^r dr \ \tanh^m (2r)=\frac{\tanh^{m+1}(2 r)}{2(m+1)}\, {}_2F_1 \big[1, \tfrac12 (1 + m), \tfrac12(3 + m), \tanh^2(2 r)\big]\ . 
\ee
For integer $m$, this reduces to simple expressions. Thus we find
\bea
Q(r)&=&\frac{1}{2}(2N_c-N_f)(2r \coth(2r)-1)\nonumber\\
&&-\frac{N_f}{2}\left(1+\sum _{k=1}^{n+1} \left(\frac{\tanh^{2k-1}(2r)}{2k}-\frac{\tanh^{k-1}(2r)}{k}-\frac{\tanh^{k+n}(2r)}{k+n+1}\right)\right)\
,\\
Q_{(n=2)}(r)&=&\frac{1}{2}(2N_c-N_f)(2r \coth(2r)-1)+\frac{N_f}{6}\tanh^2(2r)+\frac{N_f}{10}\tanh^4(2r)\
.
\label{Q2}
\eea
Notice that we have set $q_0=0$. The reason is that the term $q_0\coth(2r)$ produces a singular behavior at $r=0$, thus violating our condition that the solution reduces to
the deformed MN solution at $r=0$.

\medskip

\noindent In the following section we will proceed to the analysis of solutions $P(r)$ of the master equation (\ref{mastereq})
as a function of the parameter $x=N_f/N_c$.
In all cases we will use the $S(r)$ given by (\ref{aspa}) with $n=2$ and hence $Q$ given by (\ref{Q2}).

In general, the resulting differential equation (\ref{mastereq}) admits the following boundary conditions:
\begin{eqnarray}
P&\approx & \left\lbrace 
\begin{array}{l} 
p_0 +\mathcal O(r^3)\\
h_1 r+\mathcal O(r^3)\\
%-\frac{16}{5}r^5+\mathcal O(r^7)
\end{array}\right.
\qquad , \hspace{2.5cm} r\sim  0\ ,\\[2mm]
P&\approx & \left\lbrace 
\begin{array}{lc} 
|2N_c-N_f|\, r \ ,\ &x\neq 2\\
P_0+e^{-cr} \ ,\ &x= 2\\
k\, e^{4 r/3}   \ ,\ &{\ \rm any}\ x\ 
\end{array}\right.
\qquad ,\hspace{1.1cm}  r\gg  1\ . 
\end{eqnarray}
For each $x\neq 2$, the solution to (\ref{mastereq}), $P(r)$, is uniquely determined if we demand the following asymptotic conditions:
\begin{enumerate}
\item[a)] At $r\sim 0$, $P\sim h_1 r$, i.e. the solution reduces to the deformed MN solution with the asymptotic behavior given by (\ref{IRMMN}).
\item[b)]  At large $r$, the solution has the linear behavior (\ref{linearP}), $P\sim |2N_c-N_{f}|\, r$.
\end{enumerate}
{}For a generic integration constant $h_1$ above some critical value, the large $r$ asymptotic behavior is $P\sim e^{4r/3}$, as discussed earlier. At the critical value 
of $h_1$  the solution  has the linear behavior $P\sim(2N_c-N_{f})r$, or constant for $x=2$, and at any lower $h_1$ it meets a singularity before reaching $r=\infty$. 
Hence, the condition of  linear behavior at infinity specifies the
solution uniquely.\footnote{The solutions with exponential behavior at infinity have a constant dilaton and become Ricci flat, which is not
the expected asymptotic behavior for holographic applications. Some interesting applications of these solutions as describing properties of 6d field theories have nevertheless been found  in \cite{HoyosBadajoz:2008fw}.}

\begin{figure}[h]
\centering
\includegraphics[width=0.5\textwidth]{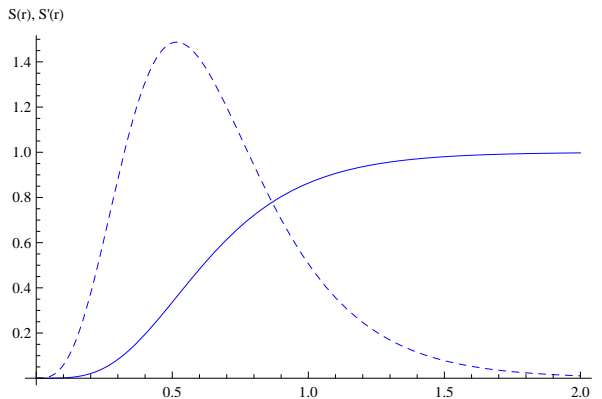}
\caption{$S(r)$ (solid line) and $S'(r)$ (dashed line).
The maximum of $S'(r)$ at $r\approx 0.5$ indicates the characteristic mass scale of the massive flavors.}\label{FSprime}
\end{figure}

\noindent In order to solve the  differential equation (\ref{mastereq}) numerically, as mentioned above we 
  take the brane distribution (\ref{aspa}) with $n=2$, and $Q$ given in (\ref{Q2}).
  This describes massive flavors with a mass around $r\approx 0.5$ (see \eqref{rmx}), determined by the maximum of $S'(r)$, shown in
  fig. \ref{FSprime} together with $S(r)$.

 \noindent Since we have to meet boundary conditions at zero and infinity,  we  employ a shooting method. 
 This determines the critical $h_1$.

%%%%%%%%%%%%%%%%%%%%%%%%%%%%%%%%%
%\subsection{$N_f<2N_c$ case} \label{sec.Nf<2Nc}
%%%%%%%%%%%%%%%%%%%%%%%%%%%%

\bigskip

\begin{figure}[tbh]
\centering
(a)\subfigure{\includegraphics[width=6.5cm]{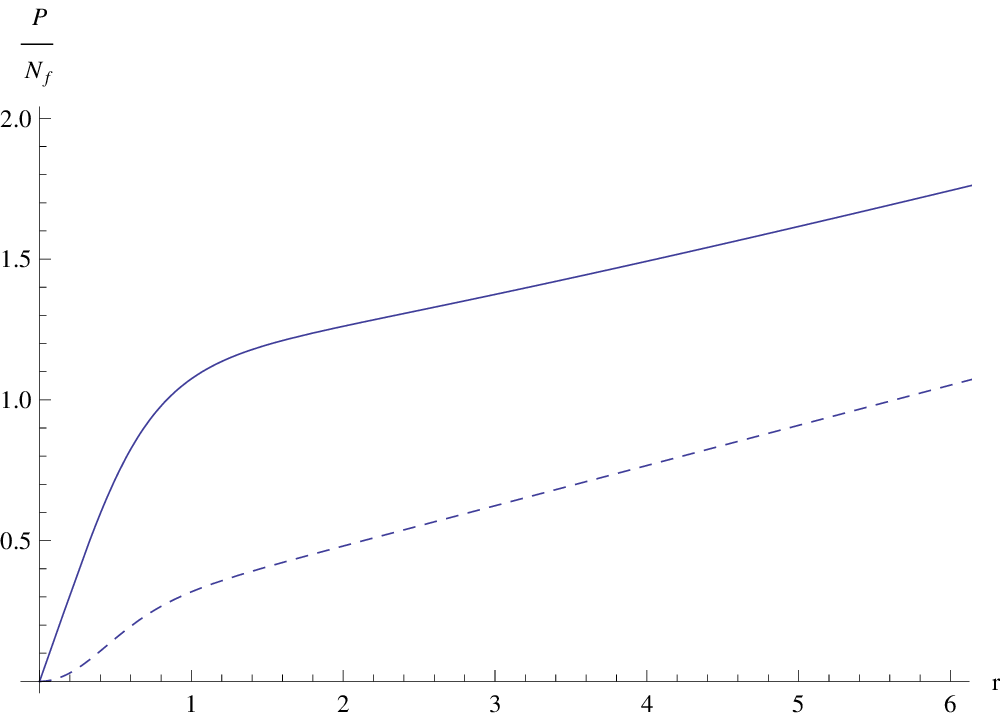}}
\ \ \ \ 
(b)\subfigure{\includegraphics[width=6.5cm]{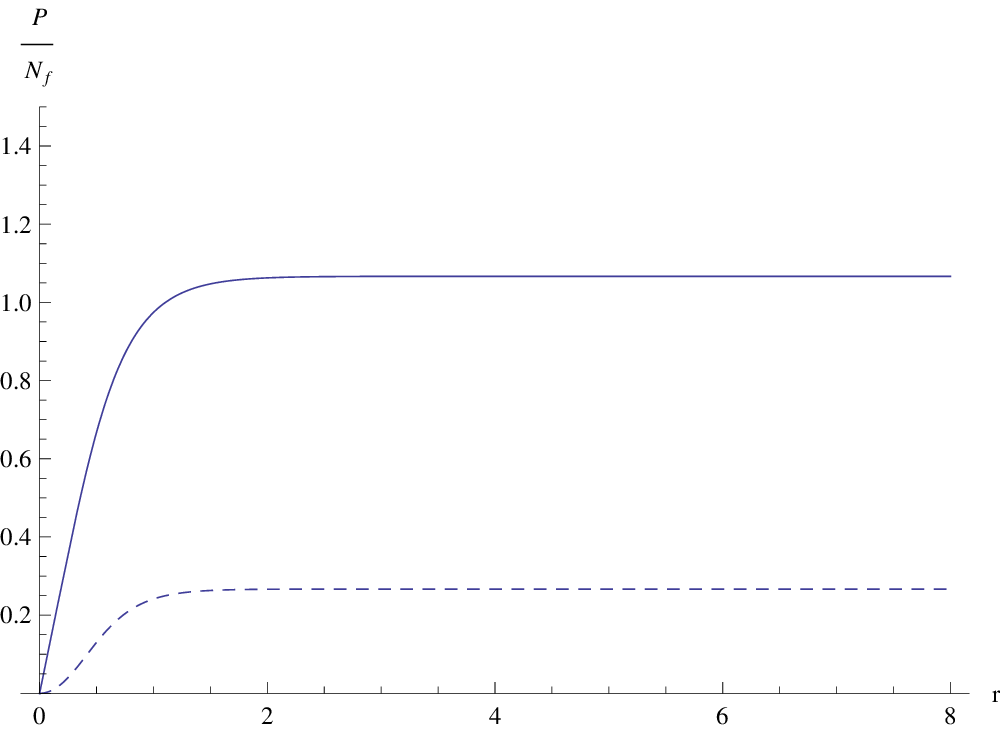}}
\\
(c)\subfigure{\includegraphics[width=6.5cm]{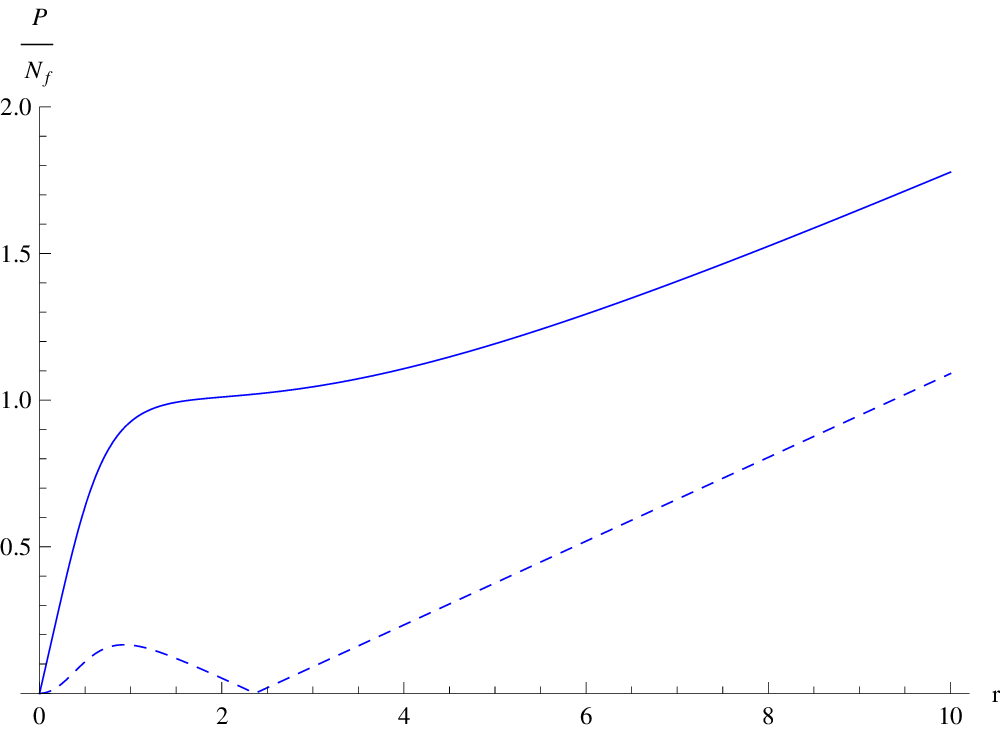}}
\ \ \ \ 
(d)\subfigure{\includegraphics[width=6.5cm]{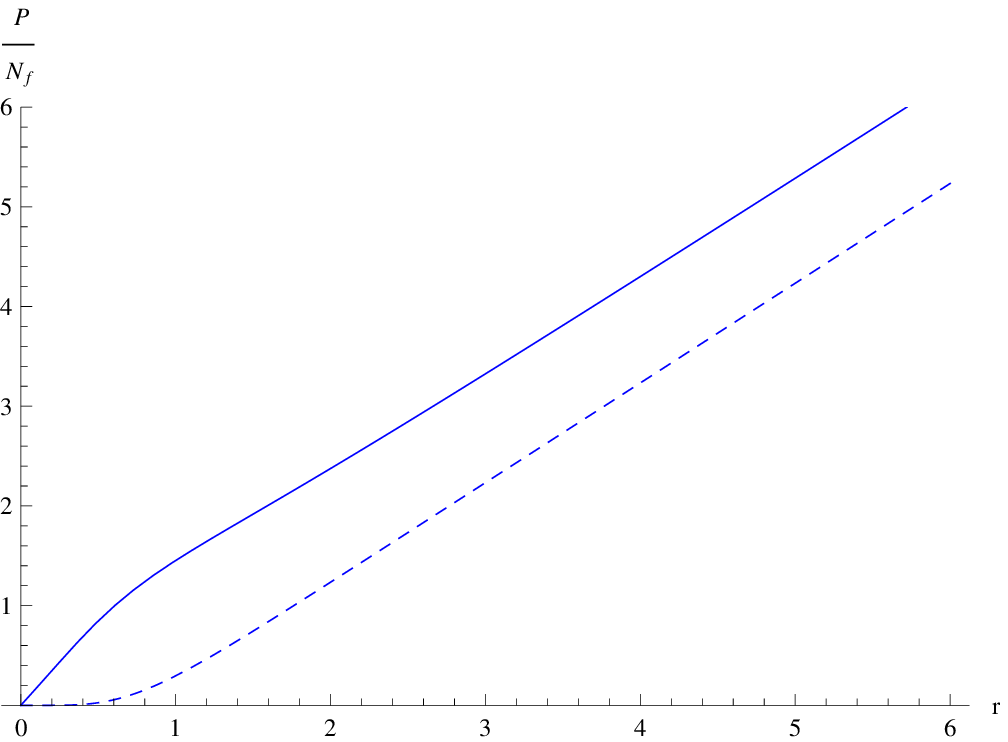}}
\caption{The function $P(r)/N_f$ solution to the master equation \eqref{mastereq}  that matches between the deformed Maldacena-N\' u\~nez solution \eqref{IRMMN} in the infrared ($r\to 0$) and the  linear behavior in the ultraviolet ($r\to\infty$). The dashed line corresponds to $Q(r)/N_f$
($|Q(r)|/N_f$ in fig. c). (a) $x=7/4$
(b) $x=2$. (c) $x=7/3$.  (d)  $x=\infty $.
\label{P-rNfless2Nc}}
\end{figure}

\noindent Figures \ref{P-rNfless2Nc}a,b,c,d  illustrate the solutions
in the three cases $N_f<2N_c$, $N_f=2N_c$ and $N_f>2N_c$.

\medskip

\begin{itemize}

\item In the first case we take $x=7/4$, for which we find
 \be
 {h_1\over N_f}\cong 1.53218706\ , \qquad x={7\over 4}\ ,\ 
 \ee
 and the solution is reported in fig. \ref{P-rNfless2Nc}a.

\medskip

\item  In the special case $x=2$ the solution that starts with $P\cong h_1 r$ near $r=0$ and asymptotes to a constant at infinity has 
 \be
 {h_1\over N_f}\cong 1.42475837\ ,\qquad x=2\ .
 \ee
 The large radius behavior is  given by
\be\label{asymptoticS=11}
P=P_0- e^{-c(r-r_1)}+\mathcal O(e^{-4r})\ ,
\ee
with
\be
 P_{0}=\frac{32N_c}{15}\ ,\qquad c=1+\frac{\sqrt{21}}{2}\ ,\qquad Q\to {8N_c\over 15}\ ,
\ee
where $r_1$ is a numerical constant. The solution is shown in  fig. \ref{P-rNfless2Nc}b.

\medskip

\item  Finally, fig. \ref{P-rNfless2Nc}c shows a case with $x>2$, taking in particular $x=7/3$, for which we find
 \be
 {h_1\over N_f}\cong 1.35890843\ ,\qquad x={7\over 3}\ .
 \ee
Note that  $x=7/3$  is related to  $x=7/4$ (used in fig.  \ref{P-rNfless2Nc}a)  by $x\to x/(x-1)$, which is produced by the change  $N_c\to N_f-N_c$.
We have made this choice for later comparison between theories related by a  naive Seiberg duality transformation.
We will comment on this below.

\end{itemize}

\medskip

\begin{figure}[h]
\centering
\includegraphics[width=0.5\textwidth]{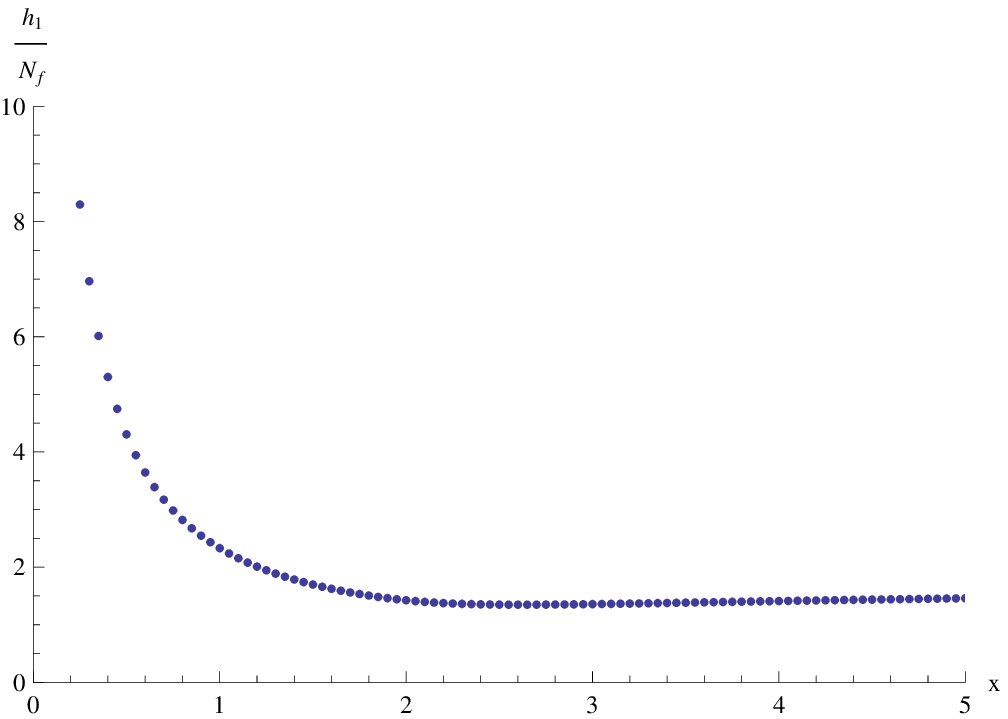}
\caption{Near $r=0$, the solutions are required to behave as $P\approx h_1 r$ to approach the deformed MN solution. 
The figure shows the critical values of the parameter $h_1$ which are required for $P$ to have
linear behavior at infinity.}\label{hhxx}
\end{figure}

\noindent More generally, one can determine $h_1$ as a function of $x$, with $0<x<\infty$, as  shown in fig. \ref{hhxx}.
For $x\to 0$ we obtain $h_1/N_f\to \infty$. Indeed one can verify that $ h_1\to 2N_f/x =2N_c$ as $N_f\to 0$, recovering the MN boundary condition at $r=0$ for $P$ discussed above. 
Furthermore, we  note that for large $x$ the critical $h_1$ approaches a finite asymptotic value, 
\be
{h_1\over N_f}\cong  1.72102763\ ,\qquad x\to\infty\ .
\ee
The reason is that for $x\gg 1$, one can scale $P\to N_f P$ so that the master equation \eqref{mastereq} becomes independent of $x$, as $Q$ becomes proportional to $N_f$, see (\ref{Q2}). This scaling solution is shown in fig. \ref{P-rNfless2Nc}d.

%%%%%%%%%%%%%%%%%%%%%%%%%%%%%%%%%%
\section{Gauge coupling $\beta$-function}
%%%%%%%%%%%%%%%%%%%%%%%%%%%%%%%%%%
\setcounter{equation}{0}

To compute the $\beta$ function of the gauge coupling in the dual field theory  we need  to first identify the gauge coupling constant in terms of geometrical quantities. For the ansatz \eqref{nonabmetric424}, this has been done in  \cite{Casero:2006pt}. The gauge coupling turns out to be directly related to the $P$ function as follows
\bea
\frac{8\pi^2}{g^2} &=&2\left(e^{2h} +\frac{e^{2g}}{4}(a-1)^2\right)=\tanh (r)\  P (r)\ .
\label{gauge-coupling}
\eea
The second crucial ingredient  necessary to obtain any $\beta $ function in the dual field theory is the precise relation between the radial coordinate $r$ of the supergravity background and the energy scale of the gauge theory. Relevant discussions on this point can be found in  \cite{Maldacena:2000yy,DiVecchia:2002ks,Bertolini:2002yr,Casero:2006pt,HoyosBadajoz:2008fw}.
In particular, in \cite{DiVecchia:2002ks,Bertolini:2002yr}, by adopting the energy scale
defined by the gaugino condensate -- a protected, i.e. gauge invariant operator with no anomalous dimensions in the gauge theory -- in the Maldacena-N\' u\~nez model, and by dimensional arguments, the following relation is obtained
\be
\label{divec}
\left( \Lambda\over \mu \right)^3 \sim a(r) \ ,
\ee
where $\mu$ is an  arbitrary renormalization scale at which the gaugino condensate is defined, and $\Lambda$ is the scale dynamically generated by quantum corrections in the gauge theory.
Strikingly, this relation applied to the unflavored Maldacena-N\' u\~nez background leads to a  gauge coupling $\beta $ function that reproduces the complete perturbative NSVZ formula (\ref{betagg}) for $N_f=0$. 
% Notice also that non-perturbative corrections related to exponentially suppressed terms $\exp (-r)$ are also present 
%\cite{DiVecchia:2002ks} and able to produce a form of soft-confinement $g_{SYM}\to const$ 
% in the far infrared, $r\to 0$ or equivalently $\mu\to \Lambda$. 
%This picture seems to suggest an approximate duality between the MN supergravity solution and ${\cal N}=1$ supersymmetric Yang-Mills theory.  
The relation (\ref{divec}) gives rise to the UV behavior
\be
\frac{\mu}{\Lambda} \sim e^{2r\over 3}\ ,\qquad r\gg 1\ .
\label{muscale}
\ee
In extending the relation between $\mu $ and $r $ to models with $N_f\neq 0$ massless flavors, one needs to consider a number of issues.
In particular,  interesting solutions exist with $a=b=0$ in \eqref{nonabmetric424}, so one should seek for other possible definitions of the energy scale than (\ref{divec}).
As emphasized  in \cite{Casero:2006pt,HoyosBadajoz:2008fw}, for a class of flavored ${\cal N}=1$ supersymmetric models,  the same UV relation (\ref{muscale})
arises from any of the following identifications
\be
\left( \Lambda\over \mu \right)^3 \sim a(r)\ ,\qquad
 \left( \Lambda\over \mu \right)^3 \sim b(r) \ ,\qquad
\left( \Lambda\over \mu \right)^3 \sim e^{-2\phi (r)}\ .
\label{aaaz}
\ee
 The relations \eqref{muscale} and \eqref{aaaz} can be generically written in the form
%use the identification between $\rho $ and $\mu $ motivated by gaugino condensate, see (\ref{aaaz}):
\be
\left( \Lambda\over \mu \right)^3 = F(r)\ ,\qquad F(r)\to e^{-2r} \ \ {\rm for}\ \  r\to \infty\ .
\label{mu2}
\ee
Different choices of $F$ are analogous to the ambiguity that appears on the field theory side in the choice of renormalization scheme.
Using \eqref{gauge-coupling} and \eqref{mu2}, we obtain  the following expression for the $\beta $ function:
 \be\label{betatau}
 \beta_{8\pi^2/ g^2}  = -{3F\over F'} \partial_{r}( \tanh r\ P)
= -{3F\over F'\cosh^2r }\left(\sinh r\cosh r\ P'+ P\right) \, 
\ee
%{}For $Q_0=0$, (\ref{aab}) gives $a(\rho) =  {1\over \cosh 2\r }$.
and by knowing $F$ and the solution $P$, we can now compute  $\beta_{8\pi^2/ g^2} $, hence $\beta _g$.
Differences between the possible radius/energy relations in \eqref{aaaz} eventually arise in the IR. 
However, we have verified that all relations in \eqref{aaaz} lead to qualitatively similar results.
For the  calculations that follow, we will adopt the prescription \eqref{divec}. In this way, when  $N_f=0$, we recover the $\beta $ function of
the MN model (specifically, the $\beta $ function obtained in \cite{Bertolini:2002yr}). 

%which differs from \cite{DiVecchia:2002ks} in subleading terms due
%to a different identification of the gauge coupling).
%which is the same adopted in the analysis of the unflavored $N_f=0$ gauge theory.

\medskip

 It is convenient to rescale away the parameter $N_f$ in  the master equation
\eqref{mastereq} by the change $P=N_f \tilde P$ and $Q=N_f \tilde Q$. This leads to
 the following scaling for the $\beta$ function,
\be
\beta_g=\frac{1}{\sqrt{N_f}}\beta_{\tilde g}(x,\tilde g)\ , \hspace{2cm} g=\frac{1}{\sqrt{N_f}}\tilde
g\ ,
\ee 
where
\be
\frac{8\pi^2}{\tilde g^2} = \tanh(r) \, \tilde P(r)\ .
\ee
In what follows we will thus compute $ \beta_ {\tilde g}$.
Note that, in terms of the  't Hooft coupling $\lambda\equiv g^2 N_c$, one has $\tilde g^2= x\lambda $, $x=N_f/N_c$, 
and 
\be
\beta_\lambda=f(x,\lambda)\ .
\label{belam}
\ee
This can be compared with the NSVZ $\beta $ function \eqref{betagg}, which  in terms of  $\lambda $ reads
\be\label{NVSZlambda}
\beta_\lambda =-{ \lambda^2\over 8\pi^2(1-{\lambda\over 8\pi^2})} (3- x(1-\gamma_0))\ .
 \ee
 This  agrees with the structure of the holographic $\beta $ function \eqref{belam}, i.e. in the large $N_c$ limit at fixed $N_f/N_c$ it only depends on $\lambda $ and $x=N_f/N_c$.

At this point it is useful to recall some basic facts of the NSVZ $\beta $ function.
It was suggested by Seiberg \cite{Seiberg:1994pq} that a conformal window for SQCD should exist for 
${3\over 2}N_c<N_f<3N_c$,  where a family of massless SQCD theories with $N_f$ massless
flavors develop an IR fixed point at finite coupling. All flavored gauge theories in the conformal window would be deconfined and chiral symmetry restored. 
The lower end-point should be considered a lower-bound on the actual value. 
A non-trivial IR fixed point can be found if $x\equiv {N_f\over N_ c}\approx 3$ \cite{BanksZaks}. 
Indeed, using the explicit form of the one-loop anomalous dimension 
the vanishing of the $\beta $ function requires
\be
{3\over x}-1 =-\gamma_0= {1\over 8\pi^2}\, g^2N_c +\mathcal O(g^4N_c^2)\, .
\label{banks}
\ee
It is clear that this fixed point moves towards the strongly coupled region as $x$ decreases from 3 to lower values.
This assumes a small value of the anomalous dimension. As we will see below, the present holographic system, like the one of  \cite{Casero:2007jj,HoyosBadajoz:2008fw}  seems to involve large values of the anomalous dimension $\gamma_0$, in fact $\gamma_0=-1/2$ in the UV.

The calculations that follow use our specific choice for the embedding function $S(r)=\tanh^4(2r)$.
However, the structure of the fixed points seems to be a generic property of the
solutions of the master equation (\ref{mastereq}) with linear dilaton asymptotic and  any  embedding function $S$ with $S(r)\to 1 $ at infinity.
This asymptotic includes previously known solutions with massless flavors. 

The linear dilaton asymptotics of these types of backgrounds preclude the emergence of an anti de Sitter  geometry at infinity, 
which should be a more appropriate description near the UV fixed points.
Despite this fact and despite the above mentioned ambiguities in the definition of the holographic beta function,
we will find some remarkable coincidences with the expected behavior in flavored SQCD.

%%%%%%%%%%%%%%%%%%%%%%%%%%%%%%%%%%%%%%%%%%%%%%%%%%%%%%%%%%%%%
\subsection{$N_f<2N_c$}
\begin{figure}[tbh]
\centering
(a)\subfigure{\includegraphics[width=6.5cm]{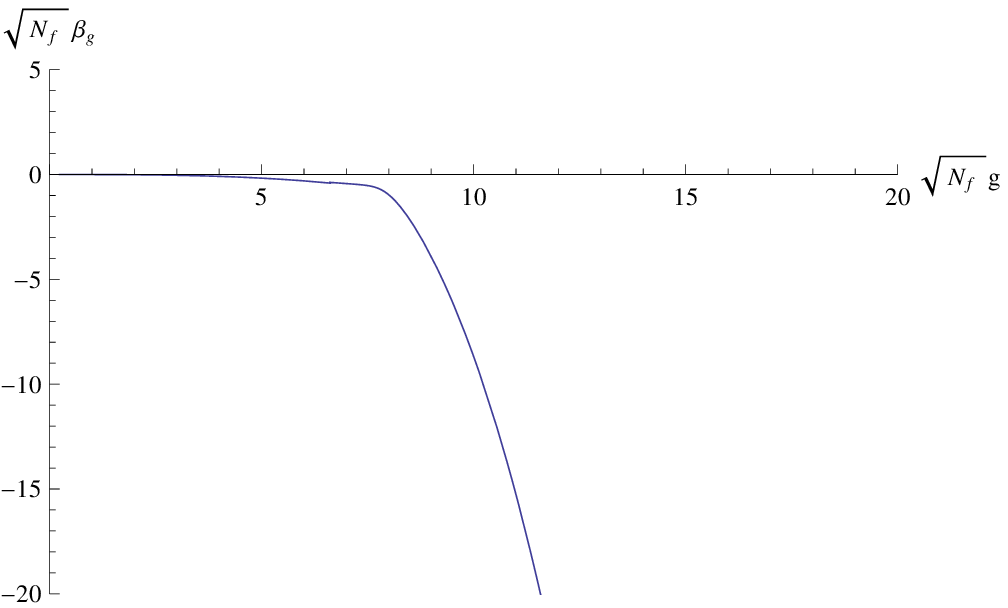}}
\ \ \ \ 
(b)\subfigure{\includegraphics[width=6.5cm]{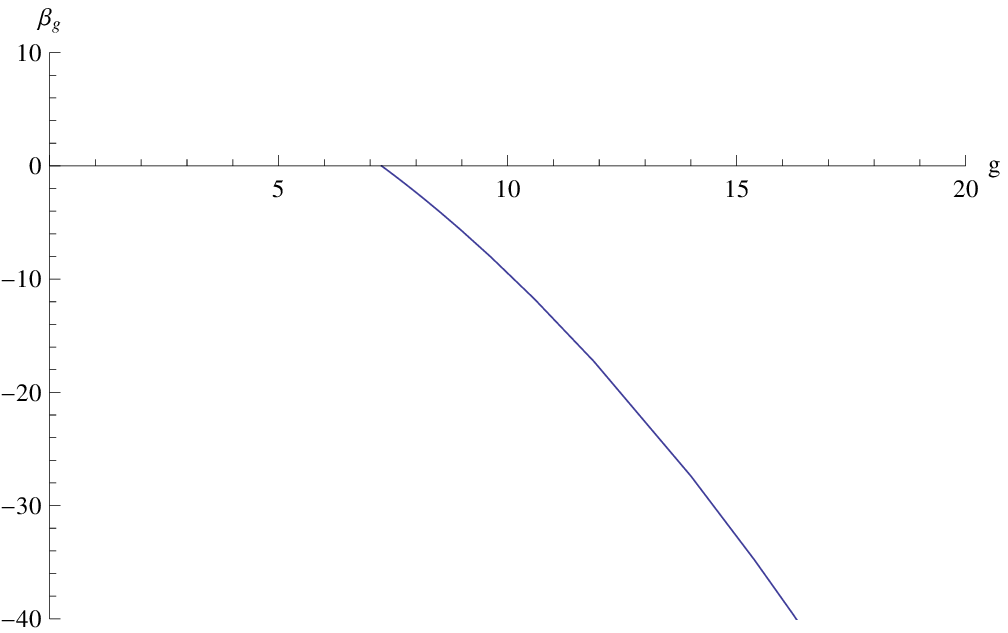}}
\\
(c)\subfigure{\includegraphics[width=6.5cm]{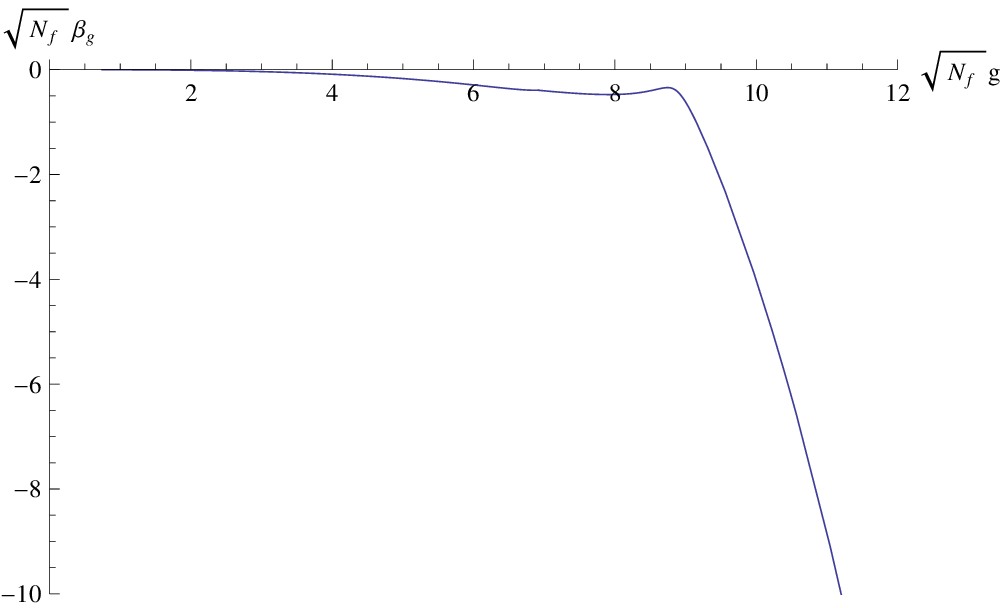}}
\ \ \ \ 
(d)\subfigure{\includegraphics[width=6.5cm]{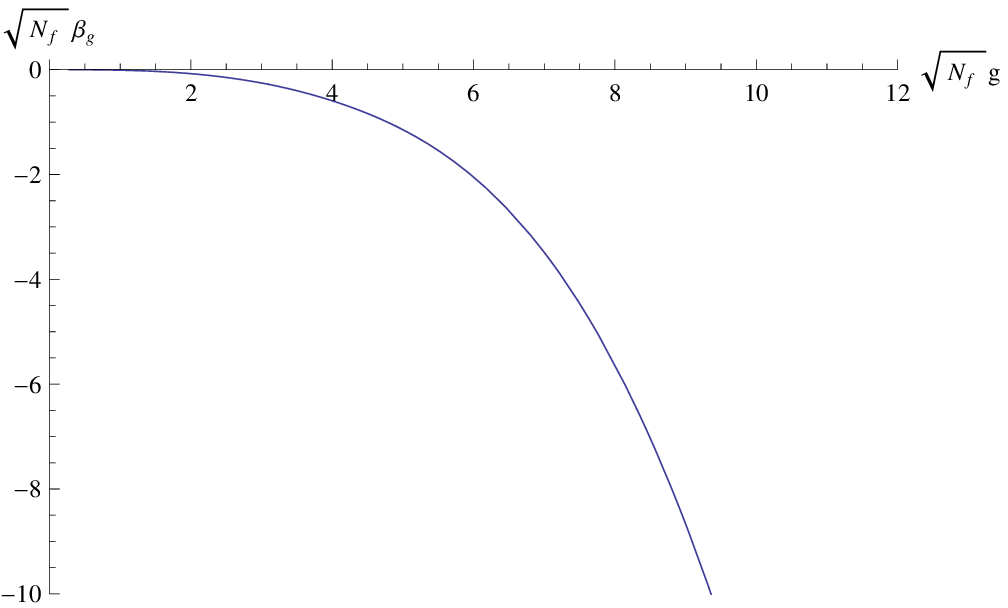}}
\caption{$\sqrt{N_f}\beta_g$ as a function of $\sqrt{N_f} g$,  corresponding to the supergravity solutions in fig.~\ref{P-rNfless2Nc}a,b,c,d.  (a)  $x=7/4$. (b) $x=2$. (c) $x=7/3$. (d) $x=\infty $.
\label{aaaaa}}
\end{figure}

The $\beta$ function for the gauge theory with massive fundamental flavors is obtained by taking the solution $P(r)$ found in the previous section (see fig. \ref{P-rNfless2Nc}a) and applying the formula \eqref{betatau}. The result is shown in fig.~\ref{aaaaa}a.

\medskip

\noindent The $\beta$ function has a UV fixed point at $g=0$, where it  has the following behavior
\begin{equation}
\beta_g\cong -\frac{3}{32\pi^2}(2N_c-N_f)g^3\, .
\label{eq:beta_less}
\end{equation}
Remarkably, this exactly agrees with the NSVZ $\beta $ function \eqref{betagg} near $g=0$, if $\gamma_0=-1/2$ in the UV --~where mass terms can be
neglected.
A similar conclusion was reached in the case of the  backgrounds with $S=1$ \cite{Casero:2007jj,HoyosBadajoz:2008fw}.
This is not surprising, since in the UV our $S$ differs from $S=1$ by exponentially suppressed terms, which do not affect the leading behavior in
\eqref{eq:beta_less}.
It would be interesting to have an independent derivation of the anomalous dimension $\gamma_0$ by holographic methods, but presently it is not clear to us what the correct prescription would be.\footnote{In \cite{Casero:2007jj} an attempt was made  to compute $\gamma_0$ by proposing that the quartic coupling of the gauge theory should be identified with some quotient of the volumes of different cycles of the manifold. }

\noindent The $\beta $ function of fig. \ref{aaaaa}a is zero at $g=0$, negative and monotonically decreasing for $g>0$, thus implying asymptotic freedom and ordinary confinement in the IR, where $g\to\infty$. 
In particular, we find no additional IR or UV fixed points at finite coupling.

%%%%%%%%%%%%%%%%%%%%%%%%%%%%%%%%%%%%%%%%%%%%%%%%%%%%%%
\subsection{$N_f=2N_c$}
Using the solution $P(r)$ found in the previous section (see fig. \ref{P-rNfless2Nc}b) we determine the $\beta $ function, shown in fig. \ref{aaaaa}b.
We can see that a non-trivial UV fixed point $g=g_*$ appears. 
%in the region where the massive flavor background converges to the massless flavor background ($S\to 1$). 
Although we do not have the gravity solution that describes the missing branch $g<g_*$, some interesting features can be inferred by comparing 
with the NSVZ $\beta $ function \eqref{betagg} in this UV region where mass terms can be neglected.
For $N_f=2N_c$, the NSVZ $\beta $ function becomes
\be
\beta_g =-{ g^3N_c\over 16\pi^2(1-{g^2N_c\over 8\pi^2})} (1+2\gamma_0 )\ .
\label{betanc}
 \ee
Again, it is consistent with our results if $\gamma_0\to -1/2$ in the UV and $g$ flows to $g_*$.
Moreover, since the  perturbative NSVZ $\beta $ function is negative near the  UV fixed point at $g=0$, by continuity there must be at least another  point $g_*'$, with $g_*'<g_*$, 
where the $\beta$ function vanishes. In the simplest assumption  that there is only one such point, this would be 
an IR fixed point. The resulting picture is in fact similar to the one proposed by Seiberg 
(for a discussion on the effect of mass terms  see e.g. \cite{Strassler:2005qs}). 
Obviously, a description using massive flavors like the present one cannot describe the emergence of a conformal fixed point in the infrared.
However, given the presence of the UV fixed point at $g=g_*$, the IR fixed point seems to be the simplest possibility that permits a negative beta function near $g=0$.
The combined presence of a pair of IR and UV fixed points is also a prerequisite for the existence of a mechanism in which the disappearance of the conformal window is due to the annihilation of a pair of fixed points \cite{Kaplan:2009kr}.
%
% To further investigate this mechanism one should  know the flavor dependence of the position of the IR and UV fixed points. {\bf Can we derive this info for the UVFP? e.g. for $N_f=2N_c\pm \epsilon$ with $\epsilon\to 0$.} 
Notice that if for $N_f=2N_c$ the IR fixed point appears at `t Hooft  coupling $\lambda=\mathcal O(1) $ (as suggested by a naive extrapolation of \eqref{banks}), it would be very difficult to see it by means of perturbative and holographic techniques.

\subsection{$N_f>2N_c$}
Using now the solution $P(r)$ of fig. \ref{P-rNfless2Nc}c we determine the $\beta $ function for the case $x=7/3$. This is shown in fig. \ref{aaaaa}c.
The $\beta$ function has, like in the $N_f<2N_c$ case, a UV fixed point at $g=0$, where  it  has the  behavior
\begin{equation}
\beta_g\cong -\frac{3}{32\pi^3}(N_f-2N_c)g^3\, .
\label{eq:betamore}
\end{equation}
This  exactly agrees with the NSVZ $\beta $ function \eqref{betagg} of the {\it  Seiberg dual gauge theory} with $\tilde N_c=N_f-N_c$ near $g=0$, if again we set $\gamma_0=-1/2$ in the UV.
This strongly suggests that in the UV region the background obtained with  our boundary conditions describes, when $N_f>2N_c$,   the Seiberg dual system.

It must be stressed that in the present case Seiberg duality is only an approximate relation that depends on the scale of energy (see \cite{Ramallo}).
The idea is that at a given scale $\mu$ one can integrate out massive flavors which have mass greater than $\mu$ and remain with a reduced number
of light flavors. 
In the present framework, this reduced number of flavors at an energy scale $r$ is effectively described by $N_f(r)\equiv N_f \, S(r)$.
With our choice of $S(r)$, massive flavors are accumulated near $r\approx 0.5$ (see fig. \ref{FSprime}).
In the infrared region, where $r\sim 0$, one has $S\sim r^4$ so $N_f(r)\to 0 $, as expected since in this region 
the energy scale is much smaller than the characteristic 
mass of the flavors. On the other hand, in the UV region, $S\to 1$ and $N_f(r)\to N_f$, which is consistent with the fact that at this scale of energies all flavors look massless.
As observed in \cite{Ramallo}, the master equation \eqref{mastereq} remains invariant under
$N_c\to N_f(r)-N_c$ and $N_f(r)\to N_f(r)$. This transformation changes  $Q(r)\to-Q(r)$. 
This is the only sense in which Seiberg duality can be applied
to the present system (in particular, $N_c\to N_f-N_c$ and $N_f\to N_f$ is not a symmetry of the master equation) and it is consistent with our proposal that the solution $P(r)$ of fig. \ref{P-rNfless2Nc}c describes the Seiberg dual system at an energy scale much larger than the characteristic mass of the flavors, where $N_f(r)\to N_f$.

Having obtained a gravity solution for the ``Seiberg dual" system, the question is how to identify a background dual to  the original gauge theory.
%We believe that the difficulties in finding this background resides in the following.
When $N_f>2N_c$, we expect that the gauge theory will develop a Landau pole. This means that the theory cannot be extended
beyond a certain UV scale. On the gravity side, it means that the geometry should terminate at a maximum value of $r$, where it probably has a singularity.
Indeed, there is a one-parameter family of solutions with parameter $h_1$ that at $r=0$ approach the deformed MN solution but at some finite $r$
meet a singularity where $P=|Q|$. These are the solutions which have an $h_1$ whose value is anything lower than the critical $h_1$ of fig. \ref{hhxx}.
In this case we lack a clear criterium to pick a unique solution in this family that is dual to the original gauge theory.
It should also be noted that the application of holography is difficult to justify for singular backgrounds that do not get to infinity.

The $\beta $ function in fig. \ref{aaaaa}c exhibits a local maximum precisely near the $g_*$ where a fixed point appears in the $N_f=2N_c$ case.
Indeed, as $N_f$ approaches $2N_c$, the local maximum approaches the line $\beta_g =0 $ and occurs at large values of $r$. 
In the strict $N_f=2N_c$ limit,
the branch $g<g_*$ disappears from the figure, because the solution gets to $r=\infty $ already at $g=g_*$.

Finally, fig. \ref{aaaaa}d shows the gauge coupling $\beta $ function computed in the infinite flavor limit, that is,
for the solution shown in fig. \ref{P-rNfless2Nc}d.
It shares similar  features as the case $x=7/3$, except that the local maximum has disappeared. 
The disappearance of the local maximum  can be understood as follows: for $x=\infty$, 
one has $\tilde x=N_f/\tilde N_c \to 1$, where $\tilde N_c=N_f-N_c$.
Thus one is computing the $\beta $ function of the ``Seiberg dual" system with $x=1$.
For $x=1$, the $\beta_g$ indeed looks very similar to fig. \ref{aaaaa}d.

%%%%%%%%%%%%%%%%%%%%%%%%%%%%%%%%%%
\section{Conclusions}
%%%%%%%%%%%%%%%%%%%%%%%%%%%%%%%%%%
\setcounter{equation}{0}

We have investigated the new gravity backgrounds found in \cite{Ramallo} dual to ${\cal N}=1$ supersymmetric gauge theories with massive fundamental flavors.
These backgrounds are characterized by a function $S(r)$ which encodes the flavor brane distribution.
In the specific backgrounds studied in this paper we have chosen a continuous $S(r)=\tanh^4(2r)$, with support in the whole space $0< r<\infty$, 
which leads to a simple analytic expression for the function $Q(r)$, and thus permits
a more straightforward integration of the master equation \eqref{mastereq} that determines $P(r)$,  hence the complete geometry.
The solutions --parametrized by $x\equiv N_f/N_c$-- were uniquely determined by imposing boundary conditions that ensure regularity at $r=0$ and
acceptable asymptotic behavior at infinity. 
In this way, solutions
are free from the  IR singularity that affects the massless flavor $S=1$ case of \cite{Casero:2007jj,HoyosBadajoz:2008fw}.  

We have then investigated
properties of the gauge coupling beta function and the possible emergence of fixed points.
As explained in \cite{Casero:2007jj,HoyosBadajoz:2008fw}, the main feature that seems to determine the properties of the 
dual gauge theory is the presence of quartic  operators in the superpotential that arise upon integration of  the Kaluza-Klein modes of the original string theory. 
These operators are of the form  $W\sim h (Q^r\tilde Q_u)(Q^u\tilde Q_r)$, with gauge indices contracted inside the parentheses and lead to
a sextic potential in the scalar fields.\footnote{ A general discussion of quartic operators can be found in \cite{Strassler:2005qs}.} They become marginal when
the anomalous dimension $\gamma_0$ is $-1/2$. For this value of the anomalous dimension the 
 NSVZ $\beta $ function \eqref{betagg} becomes
\be
\beta_g = -{3g^3\over 32\pi^2}\ {(2N_c-N_f)\over 1-{g^2 N_c\over 8\pi^2}}\, .
\label{betaNN}
\ee
We have seen that this expression agrees with the holographic $\beta $ function in the UV region for $N_f\leq 2N_c$,
and also for $N_f>2N_c$ if we replace $N_c\to N_f-N_c$. We argued that for $N_f>2N_c$ our backgrounds should therefore describe
 the Seiberg dual system, in the generalized sense discussed in sect. 4.3. After this replacement, the $\beta $ function stays negative for all 
$N_f>2N_c$. As discussed,  in the $N_f>2N_c$ case, finding  a  gravity description of the original system before Seiberg duality is difficult because,
as the expression \eqref{betaNN} indicates, asymptotic freedom is lost and thus a Landau pole is expected, presumably meaning
that the gravity solution must encounter a singularity at some radius  $r$.

For $N_f=2N_c$ we found that the theory has a UV fixed point, which hints at the presence of an IR fixed point at some lower coupling, if one is to match continuously with standard perturbative results.
In this context, we stress that our theory only converges to the massless case in the UV, and it can only asymptotically recover the presence of conformal fixed points. In particular, near the would-be IR fixed point the theory can at most have an approximate conformal symmetry at energies greater than the flavor mass scale. 
 For $N_f<2N_c$ we have not  found any evidence of an IR fixed point, perhaps suggesting  that in the presence of quartic operators the ``conformal window" opens and closes
at $N_f=2N_c$.

The effect of higher dimensional operators such as  $h (Q^r\tilde Q_u)(Q^u\tilde Q_r)$ --which in the component Lagrangian leads
to terms (scalars)$^6$ and (scalar)$^2$ (fermion)$^2$--has a counterpart in non-supersymmetric QCD.
It produces  effects that are similar to well known non-perturbative effects related to chiral dynamics in the low energy effective field theory.  For example, 
a quartic fermion operator has a key role in the emergence of chiral symmetry breaking and must have a role in the disappearance of conformality: 
 Schwinger-Dyson gap equation for the fermion propagator implies a direct relation between the onset of chiral symmetry breaking, thus the presence of a non vanishing chiral condensate,  and the point where the four-fermion operator becomes relevant in the RG flow \cite{Appelquist_ladder,Cohen}.
This happens by lowering the number of flavors in QCD-like theories, starting from the point where asymptotic freedom sets in. At sufficiently low $N_f$ chiral symmetry will always be broken and conformality
is lost. 

An interesting open problem is finding regular backgrounds that can describe the massless flavor limit in a controllable
manner. 
The current approach uses an $S(r)$ function that determined the flavor mass scale $r\sim M_f$. One can attempt to study the limit $M_f\to 0$ within this context.
Although this approach seems to be affected by  singularity problems similar to those of the massless $S=1$ case,  it is possible that some universal properties can be learned by studying this limit in detail.

%%%%%%%%%%%%%%%%%%%%
\bigskip
\bigskip

%%%%%%%%%%%%%%%%%%%%%%%%%%%%%%%%%%%%%%%%%%%%%%%%%
\section*{Acknowledgments}
%%%%%%%%%%%%%%%%%%%%%%%%%%%%%%%%%%%%%%%%%%%%%%%%%%

We would like to thank E. Conde, J. Gaillard and A. Ramallo for sending us an early version of \cite{Ramallo} prior to publication.
We are also especially grateful to A. Paredes for many useful suggestions and valuable discussions.
J.R.~acknowledges support by MCYT Research
Grant No.  FPA 2010-20807-C02-01 and project 2009SGR502. A.B. is supported by a Spanish FPU fellowship.

%%%%%%%%%%%%%%%%%%%%%%%%%%%%%%%%%%%%%%%%%%%%%%%%%


\begin{thebibliography}{20}




%\cite{Maldacena:2000yy}
\bibitem{Maldacena:2000yy}
  J.~M.~Maldacena and C.~Nunez,
  {\it Towards the large N limit of pure N = 1 super Yang Mills},
  Phys.\ Rev.\ Lett.\  {\bf 86}, 588 (2001);
  hep-th/0008001.
  %%CITATION = HEP-TH 0008001;%%

%\cite{Chamseddine:1997nm}
\bibitem{Chamseddine:1997nm}
  A.~H.~Chamseddine and M.~S.~Volkov,
  {\it Non-Abelian BPS monopoles in N = 4 gauged supergravity},
  Phys.\ Rev.\ Lett.\  {\bf 79}, 3343 (1997);
  hep-th/9707176.
  %%CITATION = HEP-TH 9707176;%%

%\cite{Karch:2002sh}
\bibitem{Karch:2002sh}
  A.~Karch and E.~Katz,
  ``Adding flavor to AdS / CFT,''
  JHEP {\bf 0206}, 043 (2002)
  [arXiv:hep-th/0205236].
  %%CITATION = JHEPA,0206,043;%%




%\cite{Casero:2006pt}
\bibitem{Casero:2006pt}
  R.~Casero, C.~Nunez and A.~Paredes,
  ``Towards the string dual of $N = 1$ SQCD-like theories,''
  Phys.\ Rev.\  D {\bf 73}, 086005 (2006)
  [arXiv:hep-th/0602027].
  %%CITATION = PHRVA,D73,086005;%%

%\cite{Casero:2006pt,Casero:2007jj,Caceres:2007mu,HoyosBadajoz:2008fw}

 %\cite{Casero:2007jj}
\bibitem{Casero:2007jj}
  R.~Casero, C.~Nunez and A.~Paredes,
  ``Elaborations on the String Dual to N=1 SQCD,''
  Phys.\ Rev.\  D {\bf 77}, 046003 (2008)
  [arXiv:0709.3421 [hep-th]].
  %%CITATION = PHRVA,D77,046003;%%




%\cite{HoyosBadajoz:2008fw}
\bibitem{HoyosBadajoz:2008fw}
  C.~Hoyos-Badajoz, C.~Nunez and I.~Papadimitriou,
  ``Remarks on the String dual to N=1 SQCD,''
  Phys.\ Rev.\  D {\bf 78}, 086005 (2008)
  [arXiv:0807.3039 [hep-th]].
  %%CITATION = PHRVA,D78,086005;%%

%\cite{Paredes:2006wb}
\bibitem{Paredes:2006wb}
  A.~Paredes,
  ``On unquenched N=2 holographic flavor,''
  JHEP {\bf 0612}, 032 (2006)
  [arXiv:hep-th/0610270].
  %%CITATION = JHEPA,0612,032;%%



%\cite{Bertoldi:2007sf}
\bibitem{Bertoldi:2007sf}
  G.~Bertoldi, F.~Bigazzi, A.~L.~Cotrone and J.~D.~Edelstein,
  ``Holography and unquenched quark-gluon plasmas,''
  Phys.\ Rev.\  D {\bf 76}, 065007 (2007)
  [arXiv:hep-th/0702225].
  %%CITATION = PHRVA,D76,065007;%%


%\cite{Caceres:2007mu}
\bibitem{Caceres:2007mu}
  E.~Caceres, R.~Flauger, M.~Ihl and T.~Wrase,
  ``New Supergravity Backgrounds Dual to N=1 SQCD-like Theories with
  $N_f=2N_c$,''
  JHEP {\bf 0803}, 020 (2008)
  [arXiv:0711.4878 [hep-th]].
  %%CITATION = JHEPA,0803,020;%%


\bibitem{Ramallo}
E.~Conde, J.~Gaillard and A.~V.~Ramallo, 
``On the holographic dual of N=1 SQCD with massive flavors",  arXiv:1107.3803.  


%%
\bibitem{Appelquist_ladder}
T.  Appelquist, K. Lane, U. Mahanta, Phys.\ Rev.\ Lett.\  {\bf 61} (1988) 1553;
T.  Appelquist, A. G. Cohen, M.  Schmaltz,  Phys.\ Rev.\ {\bf D60} (1999) 
045003.
%%

%%
\bibitem{BraunGies}
J. Braun, H. Gies, JHEP 024 (2006); J. Braun, H. Gies, arXiv:0912.4168; 
J. Braun, H. Gies,     ADD
%%

%%
\bibitem{Sannino_Susy}
T.A. Ryttov, F. Sannino, Phys.\ Rev.\ {\bf D78} (2008) 065001.
%%


%\cite{Poppitz:2009tw}
\bibitem{Poppitz}
  E.~Poppitz and M.~Unsal,
  ``Conformality or confinement (II): One-flavor CFTs and mixed-representation
  QCD,''
  JHEP {\bf 0912}, 011 (2009)
  [arXiv:0910.1245 [hep-th]].
  %%CITATION = JHEPA,0912,011;%%


%%
\bibitem{Appelquist_Lat}
  T.~Appelquist, G.~T.~Fleming and E.~T.~Neil,
  ``Lattice Study of the Conformal Window in QCD-like Theories,''
  Phys.\ Rev.\ Lett.\  {\bf 100} (2008) 171607; Phys.\ Rev.\ {\bf D79} (2009) 076010. 
%%
%%\bibitem{Hasenfratz}
A. Hasenfratz, arXiv:0911.0646; arXiv:0907.0919. 
%%

\bibitem{Hasenfratz}
  A.~Hasenfratz,
  %``Investigating the critical properties of beyond-QCD theories using Monte
  %Carlo Renormalization Group matching,''
  Phys.\ Rev.\  D {\bf 80}, 034505 (2009)
  [arXiv:0907.0919 [hep-lat]];
  %%CITATION = PHRVA,D80,034505;%%
  %``Scaling properties of many-fermion systems from MCRG studies,''
  PoS {\bf LAT2009}, 052 (2009)
  [arXiv:0911.0646 [hep-lat]].
  %%CITATION = POSCI,LAT2009,052;%%

%%
\bibitem{Deuzeman}
A.~Deuzeman, M.~P.~Lombardo, E.~Pallante, Phys.\ Rev.\ {\bf D82} (2010) 074503 arXiv:0904.4662 [hep-ph];  Phys.\ Lett.\  B {\bf 670} (2008) 41.
%%

\bibitem{kuti}
  Z.~Fodor, K.~Holland, J.~Kuti, D.~Nogradi and C.~Schroeder,
  %``Twelve massless flavors and three colors below the conformal window,''
  arXiv:1104.3124 [hep-lat].
  %%CITATION = ARXIV:1104.3124;%%

%\cite{Armoni:2009jn}
\bibitem{Armoni:2009jn}
  A.~Armoni,
  ``The Conformal Window from the Worldline Formalism,''
  Nucl.\ Phys.\  B {\bf 826}, 328 (2010)
  [arXiv:0907.4091 [hep-ph]].
  %%CITATION = NUPHA,B826,328;%%




%\cite{Novikov:1983uc}
\bibitem{Novikov:1983uc}
  V.~A.~Novikov, M.~A.~Shifman, A.~I.~Vainshtein and V.~I.~Zakharov,
  ``Exact Gell-Mann-Low Function Of Supersymmetric Yang-Mills Theories From
  Instanton Calculus,''
  Nucl.\ Phys.\  B {\bf 229}, 381 (1983).
  %%CITATION = NUPHA,B229,381;%%



%\cite{DiVecchia:2002ks}
\bibitem{DiVecchia:2002ks}
  P.~Di Vecchia, A.~Lerda and P.~Merlatti,
  ``N = 1 and N = 2 super Yang-Mills theories from wrapped branes,''
  Nucl.\ Phys.\  B {\bf 646}, 43 (2002)
  [arXiv:hep-th/0205204].
  %%CITATION = NUPHA,B646,43;%%%\cite{Bertolini:2002yr}

\bibitem{Bertolini:2002yr}
  M.~Bertolini and P.~Merlatti,
  ``A note on the dual of N = 1 super Yang-Mills theory,''
  Phys.\ Lett.\  B {\bf 556}, 80 (2003)
  [arXiv:hep-th/0211142].
  %%CITATION = PHLTA,B556,80;%%


%\cite{Cohen:1988sq}
\bibitem{Cohen}
  A.~G.~Cohen and H.~Georgi,
  %``WALKING BEYOND THE RAINBOW,''
  Nucl.\ Phys.\  B {\bf 314}, 7 (1989).
  %%CITATION = NUPHA,B314,7;%%





\bibitem{BanksZaks}
  T.~Banks and A.~Zaks,
  ``On The Phase Structure Of Vector-Like Gauge Theories With Massless
  Fermions,''
  Nucl.\ Phys.\  B {\bf 196} (1982) 189.
  %%CITATION = NUPHA,B196,189;%%
%%

 
%\cite{Kaplan:2009kr}
\bibitem{Kaplan:2009kr}
  D.~B.~Kaplan, J.~W.~Lee, D.~T.~Son and M.~A.~Stephanov,
  ``Conformality Lost,''
  Phys.\ Rev.\  D {\bf 80}, 125005 (2009)
  [arXiv:0905.4752 [hep-th]].
  %%CITATION = PHRVA,D80,125005;%%





%\cite{Koerber:2007hd}
\bibitem{Koerber:2007hd}
  P.~Koerber and D.~Tsimpis,
  ``Supersymmetric sources, integrability and generalized-structure
  compactifications,''
  JHEP {\bf 0708}, 082 (2007)
  [arXiv:0706.1244 [hep-th]].
  %%CITATION = JHEPA,0708,082;%%


%\cite{Seiberg:1994pq}
\bibitem{Seiberg:1994pq}
  N.~Seiberg,
  ``Electric - magnetic duality in supersymmetric nonAbelian gauge 
theories,''
  Nucl.\ Phys.\  B {\bf 435}, 129 (1995)
  [arXiv:hep-th/9411149].
  %%CITATION = NUPHA,B435,129;%%

%\cite{Strassler:2005qs}
\bibitem{Strassler:2005qs}
  M.~J.~Strassler,
  ``The duality cascade,''
  arXiv:hep-th/0505153.
  %%CITATION = HEP-TH/0505153;%%


%%
\end{thebibliography}
\end{document}